\documentclass[%
aip,
amsmath,amssymb,
reprint,%
]{revtex4-1}
\usepackage{graphicx}
\usepackage{dcolumn}
\usepackage{mathrsfs}
\usepackage{bm}
\usepackage{lineno}
\usepackage{hyperref}
\usepackage{bbm}
\usepackage{ulem}
\usepackage{xcolor}

\begin{document}

\title{Measurement-dependent erasure of distinguishability for the observation of interference in an unbalanced SU(1,1) interferometer}

\author{Nan Huo}
 \affiliation{College of Precision Instrument and Opto-Electronics Engineering, Key Laboratory of Opto-Electronics Information Technology, Ministry of Education, Tianjin University, Tianjin 300072, P. R. China}

 \author{Liang Cui}%
\affiliation{College of Precision Instrument and Opto-Electronics Engineering, Key Laboratory of Opto-Electronics Information Technology, Ministry of Education, Tianjin University, Tianjin 300072, P. R. China}

\author{Yunxiao Zhang}
 \affiliation{College of Precision Instrument and Opto-Electronics Engineering, Key Laboratory of Opto-Electronics Information Technology, Ministry of Education, Tianjin University, Tianjin 300072, P. R. China}

\author{Wen Zhao}
 \affiliation{College of Precision Instrument and Opto-Electronics Engineering, Key Laboratory of Opto-Electronics Information Technology, Ministry of Education, Tianjin University, Tianjin 300072, P. R. China}

\author{Xueshi Guo}%
\affiliation{College of Precision Instrument and Opto-Electronics Engineering, Key Laboratory of Opto-Electronics Information Technology, Ministry of Education, Tianjin University, Tianjin 300072, P. R. China}

\author{Z. Y. Ou}
 \email{jeffou@cityu.edu.hk}
\affiliation{Department of Physics, City University of Hong Kong, 83 Tat Chee Avenue, Kowloon, Hong Kong, P. R. China}

\author{Xiaoying Li}
 \email{xiaoyingli@tju.edu.cn}
\affiliation{College of Precision Instrument and Opto-Electronics Engineering, Key Laboratory of Opto-Electronics Information Technology, Ministry of Education, Tianjin University, Tianjin 300072, P. R. China}%

\begin{abstract}
It is known that quantum interference can disappear with the mere possibility of distinguishability without actually performing the act but can be restored by erasing the path information. Common method of erasure is by conditional projective measurement known as ``quantum eraser". On the other hand, the essence of quantum interference is amplitude addition. So, direct amplitude measurement and subsequent addition can also lead to recovery of interference. In this paper, we create temporal distinguishability in an unbalanced SU(1,1) interferometer and observe no interference in the direct photodetection of the outputs. Different from the quantum eraser scheme, we recover interference via amplitude measurement by homodyne detection even with distinguishability in photon generation but no conditional projective measurement. The new mechanism of recovering interference can be applied to other the unbalanced interferometers and should have practical applications in quantum metrology and sensing.
\end{abstract}

\maketitle

\section{Introduction}

It is well-known that the outcome in the observation of a quantum system is highly dependent on the measurement process. Hence, whether we can observe an interference effect depends on how we make the measurement \cite{bohr,feyn}. This is well illustrated in Wheeler's delayed choice experiment \cite{whee,alley} and its variance \cite{scu}, and in the quantum eraser experiments \cite{Qeraser,Qeraser2}, where the outcome even depends how we analyze the data after the measurement \cite{zei,ou97}.

\begin{figure}[t]
	\includegraphics[width=8.0cm]{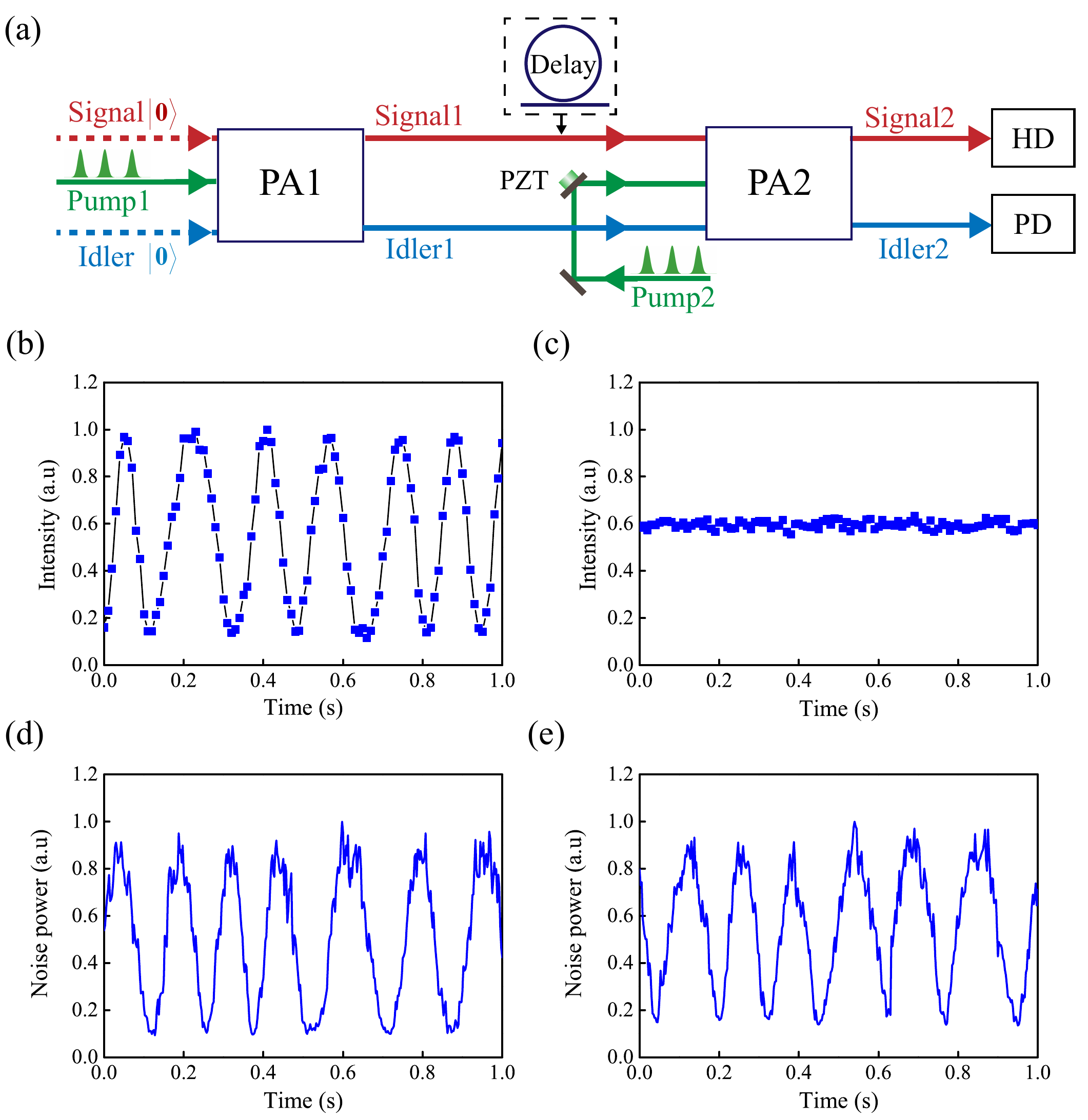}
	\caption{Observation of quantum interference in an SU(1,1) interferometer. (a) Schematics for the SU(1,1) interferometer (PZT: piezoelectric transducer for phase scan); Interference fringes in (b) direct power detection (PD) and (d) homodyne detection (HD) at the output for path-balanced interferometer; (c) Disappearance of interference in direct power detection and (e) recovery of interference fringe in homodyne detection for path-unbalanced interferometer. }
	\label{SU}
\end{figure}

Furthermore, the occurrence of quantum interference phenomena relies on whether it is possible to distinguish the interfering paths. This quantum complementarity principle is demonstrated in the phenomenon of induced coherence without induced emission \cite{zou91,zaj} in parametric processes where one (first) of the generated twin photons can be used for path distinguishability of the other twin photon (second). It seems that the disappearance of interference is predetermined by the possibility of distinguishing the first photon  even without the actual act for distinction on the photon. Nevertheless, such distinguishability can  be erased by conditional projective measurement of the first photon to recover interference of the second photon \cite{Qeraser,Qeraser2,zei}.

On the other hand, the essence of interference is the superposition of the amplitudes of the interfering fields. If we can make an amplitude measurement of the fields and add them subsequently, this should lead to interference. This is the principal idea behind the experiment that we will describe in the following, which can recover the interference via homodyne measurement even when there is distinguishability in photon generation but no projective measurement to erase it as required by the quantum eraser schemes. The observation of interference is demonstrated in an SU(1,1) interferometer \cite{yur} shown in Fig.\ref{SU}(a), where parametric amplifiers (PA1, PA2) are used to replace beam splitters for wave splitting and mixing \cite{pl10,jing11,hud14}. Such kind of nonlinear interferometers have recently been applied in many areas such as quantum imaging and spectroscopy with undetected photons \cite{lem14,kri20,kri15,tera}, quantum state engineering \cite{che15,su19,cui20}, and quantum metrology \cite{lett,du18,Liu18}. The pump field to the parametric amplifiers is a train of pulses of equal separation for the generation of photon pairs in well defined temporal mode. The interferometer is either balanced or has one of the interfering fields delayed by whole multiple of the pulse separation. We make direct power detection (PD) of intensity and homodyne detection (HD) of amplitude at the two outputs of the interferometer (after PA2) simultaneously. When the interferometer's paths are balanced, interference fringe is observed in both intensity and homodyne detection, as shown in Fig.\ref{SU}(b),(d). However, when the path difference is whole multiple of the pulse separation ($\sim$ 20 ns), interference does not occur in the direct power measurement, as shown in Fig.\ref{SU}(c) but shows up in homodyne detection, as shown in Fig.\ref{SU}(e).

The results of direct intensity measurement in Fig.\ref{SU}(b), (c)  are straightforward to understand as follows. We notice that the SU(1,1) interferometer shown in Fig.\ref{SU}(a) has similar arrangement as the frustrated two-photon interferometer \cite{herz}, which is a variation of the experiment of induced coherence without induced emission \cite{zou91,zaj}. The appearance of interference in idler (signal) side output, as shown in the observation results in Fig.\ref{SU}(b), is a result of indistinguishability \cite{zou91,zaj} of the signal (idler) photon generated from either  the first (PA1) or the second (PA2) parametric processes when the paths are balanced.  However, when a delay in the signal or idler field is introduced in the interferometer, since the generation of signal and idler photon pair is time-gated by the pump pulse, this makes it possible to distinguish which parametric process produces the signal or idler photon. This will wipe out interference in the other field (idler or signal), as shown in Fig.\ref{SU}(c) in the direct intensity measurement. The result of homodyne measurement in the balanced case shown in Fig.\ref{SU}(d) is also straightforward from the perspective of the SU(1,1) interferometer\cite{yur,hud14}.

On the other hand, the appearance of interference in the homodyne detection at the output of the unbalanced SU(1,1) interferometer is a surprise and is counter-intuitive to the argument above about dependence of interference on photon indistinguishability since photon generation is well distinguished due to pulsed pumping in two PAs for the unbalanced case.

In this paper, we analyze homodyne detection measurement process for pulsed situation and apply the analysis to the unbalanced SU(1,1) interferometer. We find that the recovery of interference in the pulse-pumped unbalanced SU(1,1) interferometer is a result of the combined effect of the unique nature of quantum measurement of amplitudes by homodyne detection and the slow response of the detection process for erasing the temporal distinguishability. In other words, the indistinguishability is exhibited in the homodyne measurement process of amplitude through slow detection. So, interference occurs really due to indistinguishability at measurement \cite{feyn} although indistinguishability in path is often exhibited in most of previous discussions on this topic. This discovery reinforces the general statement about measurement-dependence in observation of quantum phenomena \cite{bohr} and demonstrates another mechanism of recovering interference by direct amplitude addition. We start our analysis with a description of pulsed homodyne detection process and the mode coupling in pulse-pumped parametric processes.

\begin{figure}[t]
	\includegraphics[width=6.5cm]{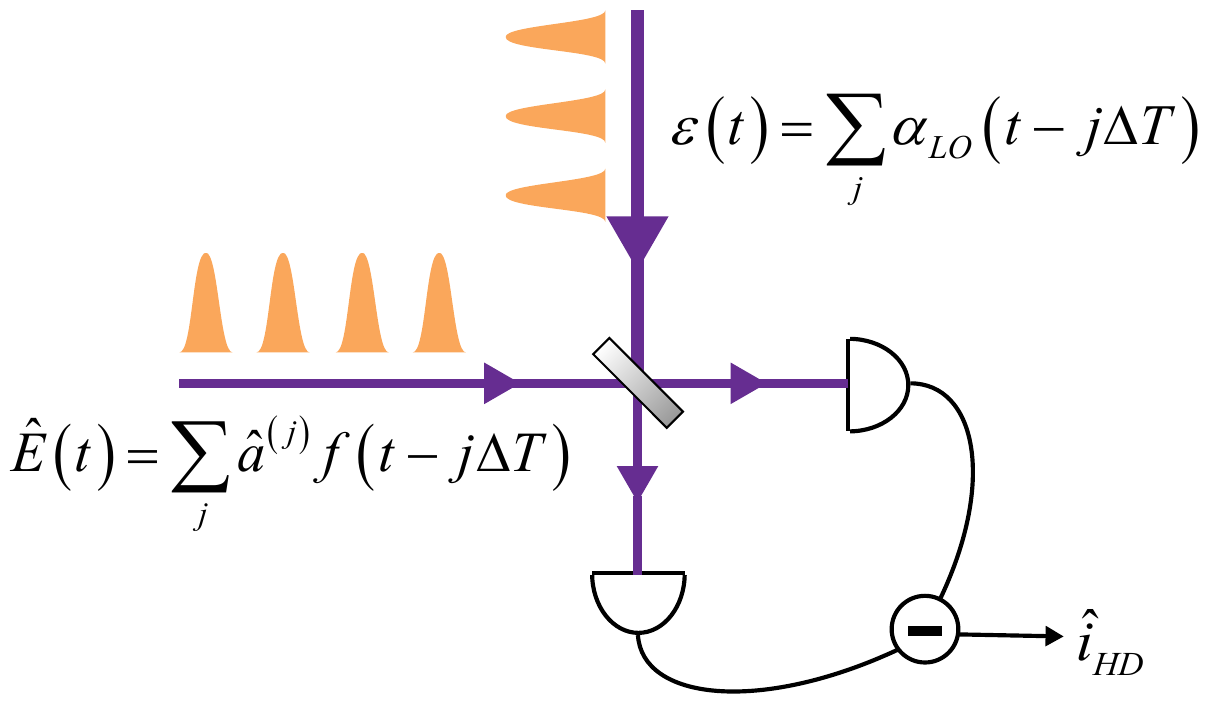}
	\caption{Schematics for homodyne detection with a train of
pulses.}
	\label{HD}
\end{figure}

\section{Homodyne detection with a local oscillator in a pulse train}

Consider the scheme shown in Fig.\ref{HD} for homodyne detection where a pulse train serves as the local oscillator (LO) which can be written as

\begin{eqnarray}\label{LO}
{\cal E}(t) &=&\sum_j \alpha_L(t-j\Delta T),
\end{eqnarray}
where $\alpha_L(t)={\cal E}_0 f(t)$ has a fixed and normalized temporal profile $f(t)$. $\Delta T$ is the separation between pulses. Assume the incoming quantum field to be detected is also described by the same temporal modes:
\begin{eqnarray}\label{QF}
{\hat E}(t) &=&\sum_j \hat a^{(j)} f(t-j\Delta T),
\end{eqnarray}
where operator $\hat a^{(j)}$ is for the $j$-th temporal mode.

We can write the output current of the homodyne detection in operator form as\cite{ou1995probability}
\begin{eqnarray}\label{iHD}
\hat i_{HD}(t) &=& \int_{T_R} d\tau k(t-\tau)\sum_j |{\cal E}_0|\hat X^{(j)}(\varphi)|f(\tau-j\Delta T)|^2\cr
&\approx&  |{\cal E}_0| \sum_j k(t-j\Delta T)\hat X^{(j)}(\varphi),
\end{eqnarray}
where $k(\tau)$ is the response function of the detector and has a width $T_R$ much larger than the pulse width of $f(t)$ so that we can pull it out of the time integral and make the approximation in the last line of the equation above. $\hat X^{(j)}(\varphi)\equiv \hat a^{(j)}e^{-i\varphi}+\hat a^{(j)\dag}e^{i\varphi}$ with $e^{i\varphi}\equiv {\cal E}_0/|{\cal E}_0|$ as the phase of the LO field.

\section{Mode coupling in parametric processes pumped by a pulse train}

Consider a train of pumping pulses described as
\begin{eqnarray}\label{Pump}
{\cal E}_p(t) &=&\sum_j E_{p0} \alpha_p(t-j\Delta T),
\end{eqnarray}
where $\alpha_p(\tau) = \int d\omega_p\alpha_p(\omega_p)e^{i\omega_p\tau}$ is the profile of the single pump pulse. Assume the pump has a Gaussian profile $\alpha_p(\omega_p)=e^{-(\omega_p-\omega_{p0})^2/2\sigma_p^2}$ with a bandwidth of $\sigma_p$.

For a single pump pulse described by $E_{p0} \alpha_p(t)$ to interact with a piece of nonlinear fiber through four-wave mixing, it is known to generate two EPR-entangled fields denoted as ``signal" and ``idler" that evolve under the unitary operator
\begin{eqnarray}\label{U}
\hat U = exp\left\{\frac{1}{i\hbar}\int dt \hat H_{eq}(t)\right\},
\end{eqnarray}
where the equivalent Hamiltonian $H_{eq}(t)$ is such that
\begin{eqnarray}
\frac{1}{i\hbar}\int dt \hat H_{eq}(t)& =& \int d\omega_1d\omega_2 F(\omega_1,\omega_2) \hat a_s(\omega_1)\hat a_i(\omega_2) + h.c. \cr &=& K\sum_k r_k \hat A_k\hat B_k + h.c.,
\end{eqnarray}
where the joint spectral function (JSF) $F(\omega_1,\omega_2)$ can be expressed by singular value decomposition as $F(\omega_1,\omega_2) = K\sum_k r_k \phi_k(\omega_1)\psi_k(\omega_2)$ with two sets of orthonormal functions $\phi_k(\omega_1),\psi_k(\omega_2)$ and normalized mode coefficients $r_k$ in descending order ($r_1>r_2>r_3>..., \sum_kr_k^2=1$). $\hat A_k\equiv \int d\omega_1 \phi_k(\omega_1)\hat a_s(\omega_1), \hat B_k \equiv \int d\omega_2 \psi_k(\omega_2)\hat a_i(\omega_2)$ define the k-th pair of temporal modes of profiles $g_k(t) \equiv \frac{1}{\sqrt{2\pi}} \int d\omega_1\phi_k(\omega_1)e^{i\omega_1 t}$ and $h_k(t) \equiv \frac{1}{\sqrt{2\pi}}\int d\omega_2\psi_k(\omega_2)e^{i\omega_2 t}$. The temporal modes are pairwise coupled by
\begin{eqnarray}\label{coupling}
\hat A_k^{out} &=& \hat A_k^{in} \cosh Kr_k + \hat B_k^{in\dag} \sinh K r_k\cr
\hat B_k^{out} &=& \hat B_k^{in} \cosh Kr_k + \hat A_k^{in\dag} \sinh K r_k.
\end{eqnarray}

The signal and idler fields can be written as
\begin{eqnarray}\label{TM}
\hat E_s^{(1)}(t) &=& \frac{1}{\sqrt{2\pi}}\int d\omega_1 \hat a_s(\omega_1)e^{-i\omega_1 t} =\sum_k g_k^*(t) \hat A_k \cr
\hat E_i^{(1)}(t) &=& \frac{1}{\sqrt{2\pi}}\int d\omega_2 \hat a_i(\omega_2)e^{-i\omega_2 t} =\sum_k h_k^*(t) \hat B_k~~~~~~
\end{eqnarray}
due to completeness relation of the mode functions: $\sum_k \phi(\omega)\phi^*(\omega')=\delta(\omega-\omega'), \sum_k \psi(\omega)\psi^*(\omega')=\delta(\omega-\omega')$. The superscript $^{(1)}$ denotes fields pumped by a single pulse.

In general, the fields are of multi-mode nature\cite{guo2015complete}. In the special case when $F(\omega_1,\omega_2)$ is factorized, the entangled fields are in single-mode. For the simplicity of discussion without loss of generality, let us assume that a single pump pulse will produce single-mode entangled signal and idler fields described by temporal mode functions $g(t), h(t)$. For a delayed pump pulse $\alpha_p(t-\Delta T)$, the pump spectrum becomes $\alpha_p(\omega_p)e^{-i\omega_p\Delta T}$ so that the JSF becomes $F(\omega_1,\omega_2)e^{-i\omega_1 \Delta T}e^{-i\omega_2 \Delta T}$. This causes the temporal modes to be also shifted by $\Delta T$ to $g(t-\Delta T), h(t-\Delta T)$. Thus, for a train of pump pulses described in Eq.(\ref{Pump}), the generated entangled fields are described by
\begin{eqnarray}
\hat E_s(t) &=& \sum_j g^*(t-j\Delta T) \hat A^{(j)} \cr
\hat E_i(t) &=& \sum_j h^*(t-j\Delta T) \hat B^{(j)},
\end{eqnarray}
where we omitted other higher-order orthogonal temporal modes in Eq.(\ref{TM}) and $\hat A^{(j)},\hat B^{(j)}$ are the operators for the corresponding pulses (temporal modes). Only operators in the same pulse ($j$) are coupled through Eq.(\ref{coupling}).

\section{Unbalanced SU(1,1) interferometer}

\begin{figure}[t]
	\includegraphics[width=7cm]{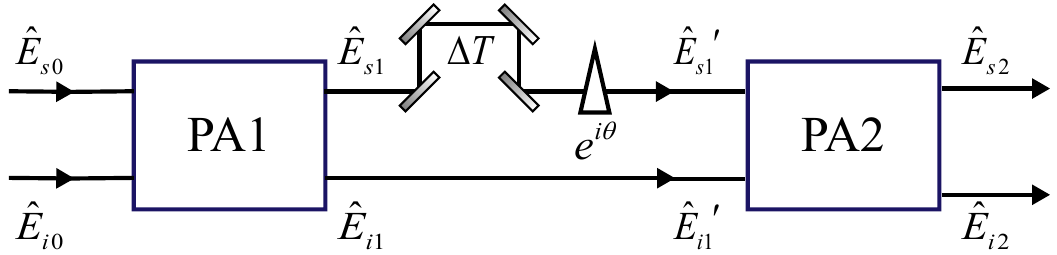}
	\caption{An unbalanced SU(1,1) interferometer with two pulse-pumped parametric amplifiers (PA1,PA2). Compared to the idler field ($\hat E_{im} (m=0,1,1',2)$), signal field ($\hat E_{sm} (m=0,1,1',2)$) has an extra delay of $\Delta T$, which is the separation between pulses. }
	\label{SUI}
\end{figure}

Referring to Fig.\ref{SUI}, the SU(1,1) interferometer with delays is formed by two pulse-pumped parametric amplifiers (PA1,PA2) which produce two coupled fields known as signal and idler. We write the corresponding fields as
\begin{eqnarray}
\hat E_{sm}(t) &=& \sum_j g^*(t-j\Delta T) \hat A_m^{(j)} \cr
\hat E_{im}(t) &=& \sum_j h^*(t-j\Delta T) \hat B_m^{(j)}
\end{eqnarray}
with $m=0$ for the initial fields, $m=1$ for the fields after the first parametric amplifier, and $m=2$ for the fields after the second parametric amplifier. Here, $\hat A_m^{(j)},\hat B_m^{(j)}$ describe the $j$-th signal and idler quantum pulsed fields in single temporal modes of $g(t-j\Delta T),h(t-j\Delta T)$, respectively. But between the two amplifiers, we introduce uneven delays for the signal and idler fields so that the input fields for the second amplifier (Fig.\ref{SUI}) are
\begin{eqnarray}\label{phase}
\hat E_{s1}'(t) &=& \sum_j g^*(t-(j+1)\Delta T) \hat A_1^{(j)} e^{i\theta}\cr &=& \sum_j h^*(t-j\Delta T) \hat A_1^{(j-1)}e^{i\theta}\cr
\hat E_{i1}'(t) &=& \sum_j h^*\big(t-j\Delta T\big) \hat B_1^{(j)} ,
\end{eqnarray}
where the primed signal field $E_{s1}^{\prime}(t)$ (Fig.\ref{SUI}) denotes the field delayed by an extra pulse separation $\Delta T$ and shifted by a phase of $\theta$. Using Eq.(\ref{coupling}) for the parametric amplifiers, we have the coupling relations for PA1
\begin{eqnarray}\label{coupling1}
\hat A_1^{(j)} &=& \hat A_0^{(j)} \cosh K_1 + \hat B_0^{(j)\dag} \sinh K_1 \cr
\hat B_1^{(j)} &=& \hat B_0^{(j)} \cosh K_1 + \hat A_0^{(j)\dag} \sinh K_1
\end{eqnarray}
and for PA2
\begin{eqnarray}\label{coupling2}
&&\hat A_2^{(j-1)} = \hat A_1^{(j-1)}e^{i\theta} \cosh K_2 + \hat B_1^{(j)\dag} \sinh K_2\cr
&&\hat B_2^{(j)} = \hat B_1^{(j)} \cosh K_2 + \hat A_1^{(j-1)\dag}e^{-i\theta} \sinh K_2
\end{eqnarray}
where for simplicity without loss of generality, we assume single mode operation for PA1, PA2  with gain parameters $K_1,K_2$. Note that because of the extra delay of $\Delta T$ in the signal field, mode $\hat B_1^{(j)}$ is now coupled to mode $\hat A_1^{(j-1)}$ of next pulse at the output of PA2. Combining Eqs.(\ref{coupling1}, \ref{coupling2}), we obtain
\begin{eqnarray}\label{coupling3}
\hat B_2^{(j)}
&=& \hat B_0^{(j)} \cosh K_1\cosh K_2 + \hat B_0^{(j-1)} e^{-i\theta} \sinh K_1 \sinh K_2  \cr &&\hskip 0.3 in + \hat A_0^{(j)\dag} \sinh K_1 \cosh K_2  \cr &&\hskip 0.5 in + \hat A_0^{(j-1)\dag} e^{-i\theta} \cosh K_1 \sinh K_2 ,
\end{eqnarray}
and similarly,
\begin{eqnarray}\label{coupling4}
\hat A_2^{(j-1)} &=& \hat A_0^{(j-1)} e^{i\theta}  \cosh K_1\cosh K_2 +  \hat A_0^{(j)}\sinh K_1 \sinh K_2  \cr &&\hskip 0.3 in + \hat B_0^{(j)\dag} \cosh K_1 \sinh K_2   \cr &&\hskip 0.5 in + \hat B_0^{(j-1)\dag} e^{i\theta} \sinh K_1 \cosh K_2 .
\end{eqnarray}

To see if there is interference, let us first evaluate the result of photo-detection at the idler output of PA2:
\begin{eqnarray}\label{iD2}
i_{i}(t) &=& \int d\tau k(t-\tau) \langle \hat E_{i2}^{\dag}(\tau) \hat E_{i2}(\tau)\rangle \cr &=&
\sum_j k(t-j\Delta T) \langle \hat B_2^{(j)\dag}\hat B_2^{(j)}\rangle  ,
\end{eqnarray}
where $k(\tau)$ is the response function of the detectors and we assume that the pulse width of $g(t)$ is much narrower than the pulse separation $\Delta T$ and the response time $T_R$ of $k(t)$ so that $\int d\tau k(t-\tau) g^*(\tau -j \Delta T)g(\tau -j' \Delta T)\approx \delta_{jj'}k(t-j\Delta T)$. Hence, averaging over the vacuum state for initial input fields $\hat A_0^{(j)},\hat B_0^{(j)}$, we have
\begin{eqnarray}\label{iD3}
i_{i}(t)&=& \sum_j k(t-j\Delta T)\Big(\langle \hat A_0^{(j)}\hat A_0^{(j)\dag}\rangle\sinh^2K_1\cosh^2K_2 \cr &&\hskip 0.5in+ \langle \hat A_0^{(j-1)}\hat A_0^{(j-1)\dag}\rangle\sinh^2K_2\cosh^2K_1\Big) \cr &=& \sum_j k(t-j\Delta T)\Big(\sinh^2K_1\cosh^2K_2 \cr &&\hskip 1.2 in+ \sinh^2K_2\cosh^2K_1\Big),
\end{eqnarray}
which shows no interference  because of photon distinguishability between different temporal modes of $\hat A_0^{(j)}$ and $\hat A_0^{(j-1)}$. We obtain similar result for direct  detection at the signal output ($i_s(t)$).

On the other hand, for the balanced case, $\hat A_0^{(j-1)}$ is replaced by $\hat A_0^{(j)}$ and $\hat B_0^{(j-1)}$ by $\hat B_0^{(j)}$ in Eqs. (\ref{coupling3}, \ref{coupling4}), which leads to interference due to indistinguishability between first two terms and between last two terms in Eqs.(\ref{coupling3}) and (\ref{coupling4}) because they both originate from the same field of $\hat A_0^{(j)}$ or $\hat B_0^{(j)}$. Or, this results in an equivalent parametric amplifier with phase-sensitive overall gain parameters of $G_T(\theta)= \cosh K_1\cosh K_2+e^{-i\theta}\sinh K_1\sinh K_2$ and $g_T(\theta) = \sinh K_1\cosh K_2+e^{-i\theta}\cosh K_1\sinh K_2$. Then, instead of Eq.(\ref{iD3}), the result of photo-detection at the idler output of PA2 becomes
\begin{eqnarray}\label{iD4}
i_{i}(t) &=& \sum_j k(t-j\Delta T) |g_T(\theta)|^2\cr &=& \sum_j k(t-j\Delta T)\big(\sinh^2K_1\cosh^2K_2 \cr &&\hskip 0.2in +\sinh^2K_2\cosh^2K_1\big) (1+{\cal V}\cos\theta)~~~~~~
\end{eqnarray}
with
\begin{eqnarray}\label{V}
{\cal V} \equiv \frac{2\sinh K_1\cosh K_2 \sinh K_2\cosh K_1} {\sinh^2K_1\cosh^2K_2 +\sinh^2K_2\cosh^2K_1},
\end{eqnarray}
which shows the interference fringe with a visibility of ${\cal V}$. Notice that the appearance or disappearance of interference is independent of the detector response $k(\tau)$ and this is consistent with experimental observation.

Next, we check the result for homodyne detection. By using Eq.(\ref{iHD}), we find  mode-matched pulsed homodyne detection of the signal field of PA2 is equivalent to the measurement of a current operator
\begin{eqnarray}\label{iHD2}
\hat i_{HDs}(t) =  |{\cal E}_0| \sum_j k(t-j\Delta T)\hat X_{A2}^{(j)}(\varphi),
\end{eqnarray}
where ${\cal E}_0\equiv |{\cal E}_0|e^{i\varphi}$ is the peak amplitude of the local oscillator pulse and $\hat X_{A2}^{(j)}(\varphi)= \hat A_2^{(j)}e^{-i\varphi}+\hat A_2^{(j)\dag}e^{i\varphi}$. Using Eq.(\ref{coupling4}) and re-organizing the indices, we have
\begin{eqnarray}\label{iHD2a}
\hat i_{HDs}(t) &=& |{\cal E}_0| \sum_j k(t-j\Delta T)\big[\hat X_{A0}^{(j)}(\varphi-\theta)\cosh K_1  \cr && \hskip 0.5in + \hat X_{B0}^{(j)}(\theta-\varphi)\sinh K_1\big]\cosh K_2 \cr\cr &&   +  |{\cal E}_0| \sum_j k(t-j\Delta T)\big[ \hat X_{A0}^{(j+1)}(\varphi)\sinh K_1 \cr && \hskip 0.5in+ \hat X_{B0}^{(j+1)}(-\varphi)\cosh K_1 \big ]\sinh K_2  \cr \cr &=& |{\cal E}_0| \sum_j k\big(t-j\Delta T\big)\big[ \hat X_{A0}^{(j)}(\varphi-\theta)\cosh K_1  \cr && \hskip 0.5in + \hat X_{B0}^{(j)}(\theta-\varphi)\sinh K_1\big]\cosh K_2  \cr\cr &&   + |{\cal E}_0| \sum_j k(t-(j-1)\Delta T)\big[\hat X_{A0}^{(j)}(\varphi)\sinh K_1 \cr && \hskip 0.5in+ \hat X_{B0}^{(j)}(-\varphi)\cosh K_1 \big ]\sinh K_2,
\end{eqnarray}
where we shift index $j$ of the second sum by 1 in the second equation. Now, since $T_R\gg \Delta T$,  we have $k\big(t-(j-1)\Delta T\big) \approx k\big(t-j\Delta T\big)$, that is, $k\big(t\big)$ changes little when shifted by $\Delta T\ll T_R$. Then Eq.(\ref{iHD2}) becomes
\begin{eqnarray}\label{iHD3}
\hat i_{HDs}(t) &\approx & |{\cal E}_0| \sum_j k\big(t-j\Delta T\big)\big\{ \big[ \hat X_{A0}^{(j)}(\varphi)\sinh K_1\sinh K_2\cr && \hskip 0.3in+  \hat X_{A0}^{(j)}(\varphi-\theta)\cosh K_1\cosh K_2\big] \cr\cr && \hskip 0.45in +\big[\hat X_{B0}^{(j)}(\theta-\varphi) \sinh K_1\cosh K_2
\cr\cr && \hskip 0.6in +\hat X_{B0}^{(j)}(-\varphi)\cosh K_1\sinh K_2 \big]\big \}.~~~~
\end{eqnarray}
Note that the expression above shows the addition of the quadrature-phase amplitude operators $\hat X$ of the same pulse ($j$), so this is equivalent to the case of no delay between the signal and idler fields in the SU(1,1) interferometer, as we discussed about Eq.(\ref{iD4}), and should give rise to interference.

To see interference more clearly, we send the output current of homodyne measurement to an electronic spectral analyzer, which measures the time average of the variance of the photo-current $(\Delta \hat i_{HDs} \equiv \hat i_{HDs} - \langle \hat i_{HDs}(t)\rangle)$:
\begin{eqnarray}
{\rm Var} = \frac{1}{T}\int_T dt \langle \Delta^2 \hat i_{HDs}(t)\rangle ,
\end{eqnarray}
where the average $\langle \rangle$ is over the quantum states of modes $\hat A_{0}^{(j)},\hat B_{0}^{(j)}$, which are all in vacuum and $T$ is the total time for the spectral analyzer to take data. With $\langle \hat X_{A0}^{(j)}(\theta)\hat X_{A0}^{(j')}(\theta')\rangle_v = \delta_{jj'} e^{i(\theta' - \theta)} = \langle \hat X_{B0}^{(j)}(\theta)\hat X_{B0}^{(j')}(\theta')\rangle_v$, $\langle \hat X_{A0}^{(j)}\hat X_{B0}^{(j')}\rangle_v = 0$ for vacuum,  a  straightforward calculation for the photo-current in Eq.(\ref{iHD2a}) gives
\begin{eqnarray}\label{S}
 {\rm Var} &=&R_p|{\cal E}_0|^2\Big\{ \int dt k^2(t)  \big[1+2\sinh^2(K_2-K_1) \cr&&\hskip 1.4in
 +  \sinh 2 K_1\sinh 2 K_2] \cr&&\hskip 0.5in+ \cos \theta
\int dt k(t)k(t-\Delta T) \cr &&\hskip 1.4in \times \sinh 2 K_1\sinh 2 K_2\Big\} \cr &=& {\rm Var}_{SN}\big[1+2\sinh^2(K_2-K_1) \cr&&\hskip 0.3in+ \sinh 2 K_1\sinh 2 K_2\big] (1 + {\cal V}_{HD} \cos \theta)
\end{eqnarray}
with visibility
\begin{eqnarray}\label{V2}
{\cal V}_{HD}(\Delta T)  &\equiv &\frac{\sinh 2 K_1\sinh 2 K_2}{1+2\sinh^2(K_2-K_1)+ \sinh 2 K_1\sinh 2 K_2}\cr&&\hskip 0.1in \times \bigg[\int dt k(t)k(t-\Delta T)\Big / \int dt k^2(t)\bigg],
\end{eqnarray}
where ${\rm Var}_{SN} \equiv R_p|{\cal E}_0|^2 \int dt k^2(t)$ is the shot noise level of HD measurement with $R_p$ as the repetition rate of the pulse train. The result in Eq.(\ref{S}) clearly shows the interference effect as we scan phase $\theta$ of the SUI, which explains the result in Fig.\ref{SU}(e), but the visibility ${\cal V}_{HD}(\Delta T) $ depends on the delay $\Delta T$ as compared to the response $k(t)$ of the HD detectors.  For $K_1, K_2\gg 1$, if $k(t)$ has a square shape of width $T_R$, for example, we have ${\cal V}_{HD}(\Delta T) = 1-|\Delta T|/T_R$ when $|\Delta T|\le T_R$ but ${\cal V}_{HD}(\Delta T) =0$ when $|\Delta T| > T_R$. The latter occurs when the detector can resolve each pulse.

The results in Eqs.(\ref{iD2},\ref{iD3},\ref{iHD2a},\ref{iHD3}) show that the difference in the observation of direct power measurement and homodyne measurement is that the former is intensity measurement which, in the unbalanced case, gives intensity addition due to distinguishability, whereas the latter is amplitude measurement which, even in the unbalanced case, results in amplitude addition from different pulses due to the slowness of detectors and thus cannot distinguish from which pulse the photon comes. In the latter case, interference occurs because of photon indistinguishability in homodyne measurement process even though there is distinguishability in photon generation in PA1 and PA2. This recovery of interference is different from the quantum eraser schemes \cite{Qeraser,Qeraser2,zei} where distinguishability in photon generation is erased by conditional projection measurement on the other correlated photon.

Notice that the result in Eq.(\ref{S}) does not depend on $\varphi$, the phase of local oscillator (LO) but only on  $\theta$, the phase of the interferometer. This is consistent with experimental observation (see next section for details) and means that the LO field does not participate in interference although it mixes with the field from the output of the interferometer. Its role is to make quadrature-phase amplitude measurement, as required to reveal the interference.

\section{Detailed Experiment}

\subsection{Unbalanced SU(1,1) interferometer with delay of multiple pulse separations}

\begin{figure*}
\centering
\includegraphics[width=14cm]{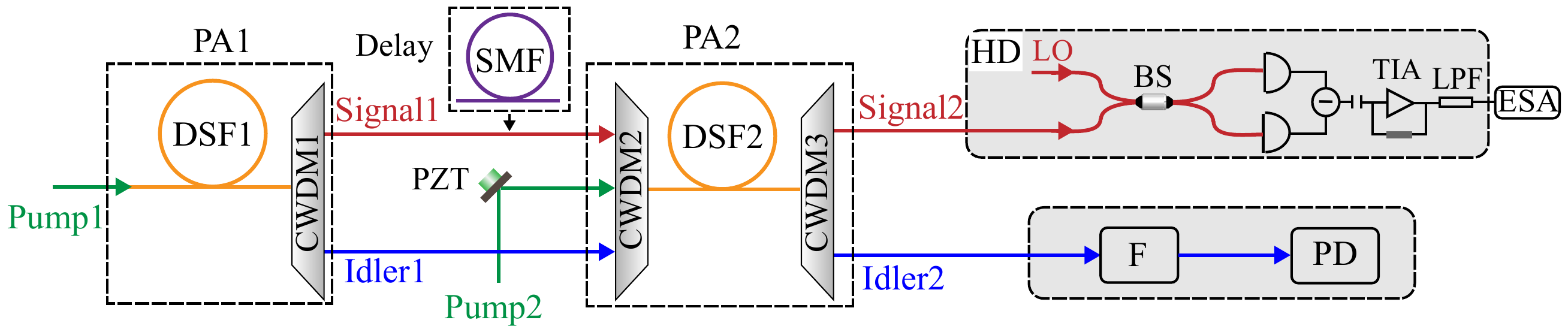}
\caption{The experimental setup for observing quantum interference in SU(1,1) interferometer, which is unbalanced when a 4.08 m-long single mode fiber (SMF) is inserted to introduce one pulse delay in the signal (or idler) channel between two PAs. PA1-2, parametric amplifier; DSF1-2, dispersion-shifted fiber; PZT, piezo-electric ceramic transducer; CWDM1-3, coarse wavelength division multiplexer; SMF, single mode fiber; F, filter; PD, power detector; HD, homodyne detection; BS, beam splitter; LO, local oscillator; TIA, transimpedance amplifier; LPF, low pass filter; ESA, electronic spectrum analyzer.}
\label{fig:Setup}
\end{figure*}

The details of our experimental setup for observing the results in Fig.\ref{SU} are shown in Fig.\ref{fig:Setup}. The SU(1,1) interferometer is formed by two PAs, which are based on four-wave mixing (FWM) in dispersion-shifted fibers (DSF). PA1 consists of a DSF and a coarse wavelength division multiplexer (CWDM); while PA2 consists of a DSF and two CWDMs. The two DSFs are identical with a zero-dispersion wavelength  of 1548.5 nm and a length of 150 m. They are submerged in liquid nitrogen (77 K) to suppress Raman scattering. The central wavelength of pulsed pumps for the two PAs (P1 and P2) is set at 1549.3 nm, so that the phase matching condition of FWM with a gain bandwidth up to 40 nm is achieved in DSFs \cite{han01,li08}. The CWDMs have three channels for pump, signal, and idler fields, respectively, with a bandwidth of 16 nm for each. They are used for the separation or combination of fields at different wavelengths.

PA1 generates the entangled signal and idler fields via spontaneous FWM, and CWDM1 is used to separate signal1 and idler1 fields from pump P1. After propagating the two outputs of PA1 along different paths, CWDM2 couples pump P2 and signal1 and idler1 fields into DSF2, and CWDM3 separates signal2 and idler2 fields from P2. The signal and idler fields at both the input and output of PA2 are broadband due to the large bandwidth of the CWDMs. Note that to ensure the modes of all the fields involved in FWM in PA2 are well overlapped, there are fiber polarization controller and adjustable delay lines in both the signal1 and idler1 channels (not shown in Fig.\ref{fig:Setup}). The procedure and method used for optimizing the gain of FWM by controlling the temporal and polarization modes are described in more detail in our previous experiment \cite{Guo16,Liu18}. Varying the phase $\phi$ of P2 by scanning a piezoelectric transducer (PZT) is equivalent to introducing a phase shift $\theta=\frac{\phi}{2}$ in Eq. (\ref{phase}) (or Fig. \ref{SUI}).

The pulsed pumps and local oscillator (LO) of homodyne detection are created by  carving the output of a mode locked fiber laser with a dual-band filter (DBF, not shown) \cite{Guo16}. The laser output is in the range of 1520-1600 nm. The pulse duration and repetition rate of the laser are 150 fs and 50 MHz, respectively.
So the time period $\Delta T$ between two adjacent pulses is 20 ns. The wavelengths of the pump and LO are set at 1549.3 nm and 1532.5 nm, respectively.
To achieve the required power, the pumps and LO carved out by DBF are respectively amplified by erbium-doped fiber amplifiers and reshaped
by the programmable filters (Waveshaper 4000A from Finisar) \cite{Liu-OL18}. The pulse durations of the transform-limited pulse trains for P1, P2, and LO are about 12, 8 and 3 ps, respectively.

At the output of PA2, the signal2 and idler2 ports are respectively analyzed by balanced homodyne detection (HD) and direct power detection (PD) at the same time. The HD at signal2 output port is formed by a 50/50 beam splitter and two photo-diodes. The difference of the AC photocurrents from two photo diodes is amplified by a transimpedance amplifier (TIA) and passes through a 1.9 MHz low-pass filter. So the response time of the HD is about 0.5 $\mu$s, which is much longer than the delay $\Delta T$ (20 ns). The direct power detection measurement is carried out at idler2 output port after passing through a filter (F) to further isolate the pump. The filter F with a 3 dB bandwidth of 1 nm is centered at 1566.5 nm. We use two kinds of detectors for direct intensity measurement: one is fast enough to resolve single pulse, while the other is a slow one that is unable to resolve each pulse. The fast PD is realized by launching the attenuated idler2 field into an InGaAs-based single photon detector, which is operated in gated Geiger mode and the gate rate is the same as the repetition rate of pump pulses.  The efficient gate width of single photon detector (about 500 ps), equivalent to the response time of the PD, is much longer than the pump pulse duration (12 ps), but is much shorter than the separation between two adjacent pump pulses, $\Delta T$ (20 ns). In the process of directly measuring the power, the integration time of single photon detector for each data point (see the solid squares in Figs. 1(b) and 1(c)) is set to 10 ms. The slow PD is realized by a detector having a response bandwidth of about 750 Hz. In our experiment, the phenomena observed by PD are not dependent upon the detection response time, which is consistent with the theory.

\begin{figure}[t]
\includegraphics[width=8cm]{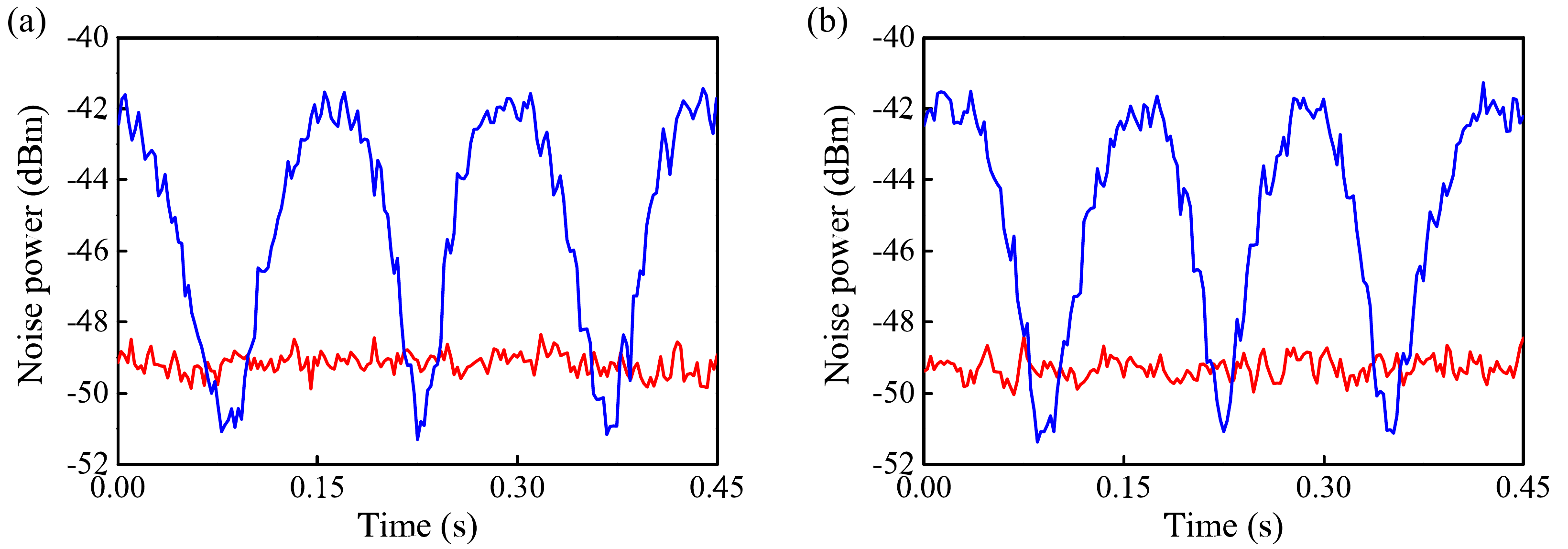}
\caption{Noise levels in log-scale measured by homodyne detection as a function of phase scan when the SU(1,1) interferometer is (a) balanced and (b) unbalanced, respectively. }
\label{noise}
\end{figure}

We first perform the measurement when the SU(1,1) interferometer is balanced. In this situation, the path lengths of signal1 and idler1 fields between two PAs are identical, the output of PA2 becomes phase sensitive and its output noise is dependent upon the phase difference between pump P2 and two input fields to PA2 \cite{JMLI-PRA20}. So the interference fringes in both intensity and homodyne measurement are expected. During the measurement, the average powers of the pumps P1 and P2 are 2.3 mW and 3.2 mW, respectively, and the power gains of PA1 and PA2 are $G_1^2={\rm cosh}^2 K_1= 2.8$ and $G_2^2={\rm cosh}^2 K_2= 10$, respectively.  This setting is to optimize the observed noise reduction (see later). In order to observe the interference effect, the phase of P2 is slowly scanned at a rate of 1 Hz by applying a ramp voltage on a piezoelectric transducer (PZT). It is clear that interference can be observed in the results of PD and HD measurement, as shown in Fig.\ref{SU}(b),(d), respectively. The visibility is about 77 $\%$ in Fig. \ref{SU}(b) and 79 $\%$ in Fig. \ref{SU}(d). The deviation of the visibility from the maximum value of 100\% is because the gains of two PAs have not been optimized \cite{NH-UP21}. However, under the current pump powers of P1 and P2, we are able to observe optimum noise reduction of two entangled fields by HD, as shown in Fig. \ref{noise}(a). The red line in Fig. \ref{noise} is obtained by blocking P1, which means the input of PA2 is vacuum and the noise measured by HD corresponds to the shot noise level (SNL) in the PA-assisted entanglement measurement scheme \cite{JMLI-PRA20}. Comparing the blue curve and SNL level (red) in log-scale in Fig. \ref{noise}(a), we find the entanglement between signal1 and idler1 leads to a noise reduction in the measured photocurrent, which is lower than the SNL by 1.5 dB.

Next, we insert a 4.08-m-long standard single-mode fiber (SMF) at signal1 of PA1, so that the SU(1,1) interferometer become unbalanced with a delay of exactly the separation between adjacent pulses. In this experiment, all the parameters are the same as the balanced case, except the optical path delay induced in signal1 field. We simultaneously carry out the direct intensity measurement and homodyne measurement at the two outputs of the unbalanced SU(1,1) interferometer, respectively. Figure \ref{SU}(c) shows that the power measured by PD stays constant, indicating there is no interference observed; while the result of HD in Fig. \ref{SU}(e) is the same as that in Fig. \ref{SU}(d), indicating the noise level at the individual output of unbalanced SU (1,1) interferometer changes with the phase of pump P2, showing the interference effect. Note that when the one pulse delay is switched to idler1 field, the results are the same as those in Figs. \ref{SU}(c) and \ref{SU}(e). The experimental results agree well with the theoretical prediction given in Eqs. (\ref{iD3}) and (\ref{S}). The noise levels in log-scale in this unbalanced case shown in Fig. \ref{noise}(b) are the same as in Fig. \ref{noise}(a). The result indicates that although PA2 does not function as a phase sensitive amplifier and is unable to coherently combine the two entangled fields generated from PA1 due to the delay induced unbalance, the low noise performance of SU(1,1) interferometer \cite{yur, JMLI-PRA20}, which is the key for its application in quantum sensing, can be retained by maintaining quantum interference via homodyne detection.

\subsection{Unbalanced SU(1,1) interferometer with arbitrary delay}

\begin{figure}[t]
	\includegraphics[width=8cm]{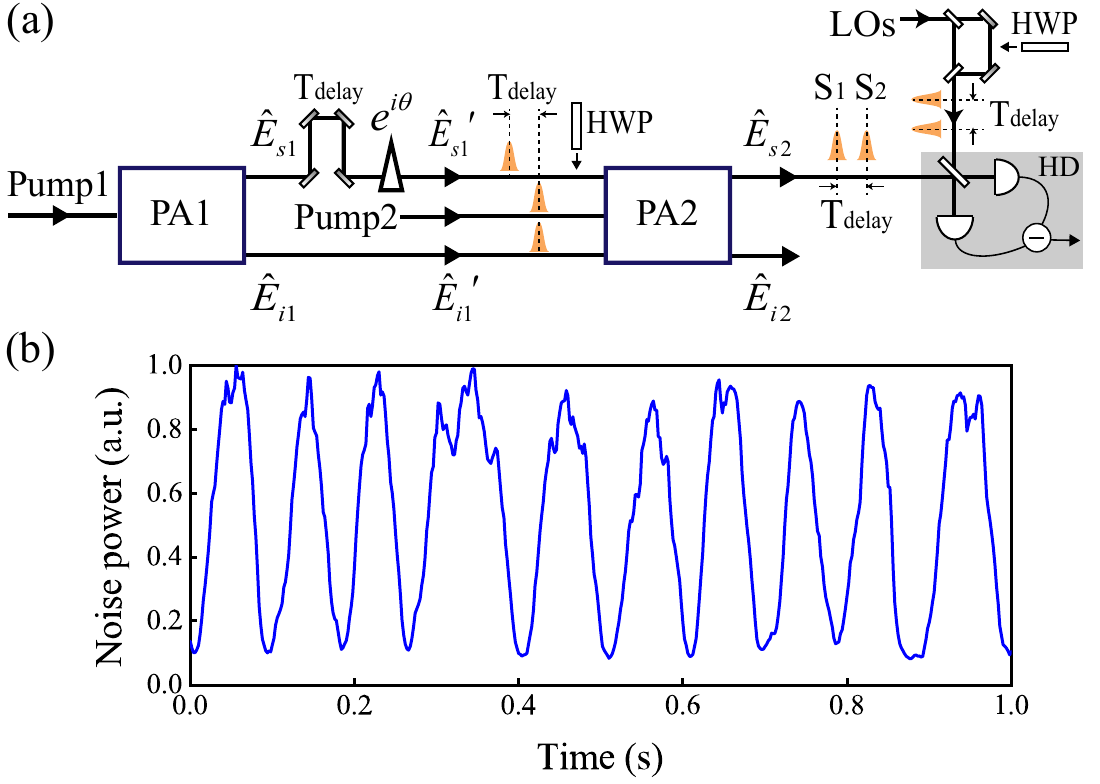}\\
	\caption{(a) An unbalanced SU(1,1) interferometer with two pulse-pumped parametric amplifiers (PA1,PA2) and asynchronized signal arm. LO consists of two pulses to measure the asynchronized signal field for amplitude addition. HWP, half-waveplate. (b) Observed interference fringe in the homodyne measurement. }
	\label{asyn}
\end{figure}

In order to prove further that amplitude addition by homodyne detection is what leads to the recovery of interference in the unbalanced SU(1,1) interferometer, we make a variation of the unbalanced interferometer by setting the delay of the signal arm not equal to a multiple of the pump pulse separation but some arbitrary number ($T_{delay}\ne N\Delta T$ in Fig.\ref{asyn}(a)). In this case, the second PA will not interact the delayed signal field generated in the first PA with the corresponding idler field at all because the pump pulse to the second PA is not synchronized with the delayed signal field. But to measure the asynchronized signal field (S1), we need to make an extra pulse in the LO so that the homodyne detection still measures the amplitude of the asynchronized signal field and adds it with the amplitude of the signal field (S2) that is generated by the second PA and is synchronous to the second pump pulse. The theory for explaining this experiment is similar to Eq.(\ref{iHD2a}) except that there is no gain from PA2 for the delayed signal ($K_2=0$ in the terms containing $X_{A0}^{(j)}$ and $X_{B0}^{(j)}$ or in the terms containing $A_0^{(j-1)}$ and $B_0^{(j-1)\dag}$ in Eq.(\ref{coupling4})).

For the arrangement in Fig.\ref{asyn}(a), we replace the 4.08 m-long fiber (corresponding to one pulse delay) with a 4.10 m-long fiber in the signal arm so that the signal field generated by PA1 is not synchronized with the pump pulse and idler field at PA2. For the homodyne measurement of the delayed signal field, we create a second pulse in local oscillator (LO) by using an unbalanced Mach-Zehnder (MZ) interferometer with the same delay. When the pump phase of PA2 is scanned, interference fringe shown in Fig.\ref{asyn}(b) is observed. In this experiment, the power gain of PA2 is set at $G_2^2 = 2$ for maximum visibility. When one arm of the unbalanced MZ interferometer is blocked, no interference is observed in HD measurement, demonstrating the role of mode selection for the LO in this experiment.

To make the interference recovered by HD measurement more dramatic, we insert a half-wave plate(HWP) in the delayed signal field and rotate its polarization by 90 degree. The corresponding LO also has its polarization rotated 90 degree by inserting a HWP in the delayed arm of MZ interferometer. In this way, there is no interference in direct intensity measurement no matter what is done but we observe the same interference fringe in homodyne measurement as Fig.\ref{asyn}(b). This further confirms the mode selectivity of homodyne detection and the amplitude addition for the recovery of interference.

\section{Discussion}

How to understand in a consistent way the disappearance of interference in direct intensity measurement and the recovery of interference by homodyne detection (HD) in the unbalanced case? At first, it seems trivial since homodyne measurement involves mixing of local oscillator (LO) with the detected fields and the result could be the interference between LO and the field. However, this is ruled out immediately because we observe in the experiment that the interference only depends on the phase of the interferometer but not on that of the LO. Secondly, slow response of detectors in HD could also be a reason since the signal measured in HD is analyzed by an electronic spectral analyzer at 1 MHz while the detector in direct intensity measurement has a response time of 0.5 ns and can resolve the pulse trains of separation of 20 ns. So we switch to a slow detector with a 3 dB bandwidth of 750 Hz for direct intensity measurement in the unbalanced case. But no interference occurs as before.
It is further speculated that narrow band detection in homodyne measurement might be the cause so it is suggested that optical filtering will be able to lengthen the coherence time of the field and allow the delayed signal field to overlap with the signal field generated in PA2. To check this, we make a variation of the unbalanced interferometer, in which the delay of the signal arm  is not equal to a multiple of the pump pulse separation and the polarization of delay signal is rotated by 90 degree. In this case, no matter how narrow the filter is, there is no way to observe interference in direct power detection. Therefore, the role played by homodyne detection is not a simple narrow band detection.

Since the explanation of the appearance of interference in homodyne detection is based on amplitude superposition, which is similar to the classical wave picture, a question thus arises about classical explanation of the observed results. Indeed, there exists a classical wave model in terms of phase-sensitive amplification \cite{vered15}. However, this model should likewise predict the appearance of interference in direct intensity measurement, as we discussed in Eq.(\ref{iD4}) for the balanced case. This is in contradiction with the observation when a delay is present. On the other hand, the photon (particle) picture in terms of distinguishability in photon generation seems to give prediction of no interference in both intensity and homodyne measurement. Therefore, this is the wave-particle duality at play again and it is exhibited in one experiment (with delay) but in different measurement schemes. The full quantum explanation we presented is required to give correct results as observed in both cases. Furthermore, the quantum feature for homodyne measurement stems from the visibility formula in Eq.(\ref{V2}), which shows the dependence on the time delay. This is because the vacuum states for the temporal modes of different pulses ($\hat A_0^{(j)}, \hat A_0^{(j-1)}$ and $\hat B_0^{(j)}, \hat B_0^{(j-1)}$ in Eqs.(\ref{coupling3},\ref{coupling4})) are different and independent, and therefore are distinguishable. But the slow detection response in homodyne detection makes them indistinguishable in amplitude, leading to interference.

\section{Summary and future perspective}

In summary, we investigated an SU(1,1) interferometer with the unbalanced paths. We observe no interference in the direct intensity measurement but demonstrate the reappearance of interference in homodyne detection. The absence of interference is because of the photon distinguishability in the delayed path but the recovery of the interference is a result of amplitude superposition in pulsed homodyne detection measurement process. Such kind of measurement-dependent observation of quantum interference demonstrates the dependence on measurement in the observation of quantum phenomena.

The unbalanced interferometers have practical implication in optical sensing where one arm of the interferometer usually goes through a long sample. If interference exists in the unbalanced case, it is not necessary to introduce path compensation in the other arm, creating flexibility and convenience in quantum sensing applications. In quantum probing by undetected photons, however, direct detection method by photon counting has so far been used \cite{lem14,kri20,kri15,tera}, which leads to no interference in the unbalanced case. Our approach with homodyne measurement will be able to overcome this huddle. Notice that because of the short pulses for pumping and the phase matching of parametric process in each PA, the signal and idler fields are broadband and suitable for spectroscopic sensing.  Concerns may arise about the requirement of mode match for homodyne measurement in both temporal and spatial degrees of freedom. However, this is not a problem here because SU(1,1) interferometer is resilient to any detection losses, to which mis-match in mode  is equivalent \cite{JMLI-OE19,JMLI-PRA20}.  Furthermore, the role of homodyne is mode selection. Large mode mismatch due to mode distortion when light passes through medium will reduce indistinguishability and thus the visibility of interference. This provides another quantity to measure for sensing, just like the technique in imaging with undetected photons \cite{lem14,kri20,kri15,tera}.

On the other hand, if detectors are fast and can resolve the fields of different pulses, distinguishability will lead to no interference in homodyne detection, as shown in Eq.(\ref{V2}).
This situation provides us further with a method to re-mix the original fields with the delayed fields, creating superposition of different temporal modes, as shown in Eqs.(\ref{coupling3},\ref{coupling4}), for the generation of cluster states \cite{clus1,clus2}.  By comparing our experimental setup in Fig. \ref{SU}(a) to the setup reported in Refs.\citenum{clus1,clus2}, one sees that the difference is that we replace the second beam splitter (BS2) in the cluster state generating setup with the parametric amplifier PA2.
In the sense of combining signal and idler beams generated by PA1, the function of PA2 is equivalent to that of BS2 in Ref. \citenum{clus1}.
Thus, it is possible to generate multi-partite entanglement state with the unbalanced SU(1,1) interferometer once the detector is fast enough, though the specific configuration of the interferometer may be slightly different from our setup.

\end{document}